\documentclass[aps, prl, preprint,superscriptaddress,groupaddress, amsmath,amssymb, footinbib, showkeys]{revtex4-1}
\usepackage{epsfig,psfrag}
\usepackage{dcolumn}
\usepackage{bm}
\usepackage{epsfig} 
\usepackage{graphicx}
\usepackage{epstopdf}
\usepackage{xcolor}
\usepackage{array}
\usepackage{import}
\usepackage{siunitx}
\usepackage{transparent}
\begin{document}
\title{Semiconductor quantum plasmonics}

\author{Angela Vasanelli}\email{angela.vasanelli@ens.fr}

\affiliation{Laboratoire de Physique de l'Ecole normale sup\'erieure, ENS, Universit\'e PSL, CNRS, Sorbonne Universit\'e, Universit\'e de Paris, Paris, France}

\author{Simon Huppert}\altaffiliation{Current affiliation: Institut des Nanosciences de Paris, CNRS, Sorbonne Universit\'e, 75005 Paris, France}
\affiliation{Laboratoire Mat\'eriaux et Ph\'enom\`enes Quantiques, CNRS - UMR7162, Universit\'e de Paris, Paris, France}

\author{Andrew Haky}

\affiliation{Laboratoire de Physique de l'Ecole normale sup\'erieure, ENS, Universit\'e PSL, CNRS, Sorbonne Universit\'e, Universit\'e de Paris, Paris, France} 

\author{Thibault Laurent}

\affiliation{Laboratoire Mat\'eriaux et Ph\'enom\`enes Quantiques, CNRS - UMR7162, Universit\'e de Paris, Paris, France}

\author{Yanko Todorov}

\affiliation{Laboratoire de Physique de l'Ecole normale sup\'erieure, ENS, Universit\'e PSL, CNRS, Sorbonne Universit\'e, Universit\'e de Paris, Paris, France} 

\author{Carlo Sirtori}

\affiliation{Laboratoire de Physique de l'Ecole normale sup\'erieure, ENS, Universit\'e PSL, CNRS, Sorbonne Universit\'e, Universit\'e de Paris, Paris, France}

\begin{abstract}
We investigate the frontier between classical and quantum plasmonics in highly doped semiconductor layers. The choice of a semiconductor platform instead of metals for our study permits an accurate description of the quantum nature of the electrons constituting the plasmonic response, which is a crucial requirement for quantum plasmonics. Our quantum model allows us to calculate the collective plasmonic resonances from the electronic states determined by an arbitrary one-dimensional potential. Our approach is corroborated with experimental spectra, realized on a single quantum well, in which higher order longitudinal plasmonic modes are present. We demonstrate that their energy depends on the plasma energy, as it is also the case for metals, but also on the size confinement of the constituent electrons. This work opens the way towards the applicability of quantum engineering techniques for semiconductor plasmonics.

\end{abstract}


\maketitle

The properties of plasmons in nanostructures can be profoundly modified whenever their elementary constituents, electrons and photons, enter the quantum regime~\cite{halperin, quantum_plasmonics, review_IEEE, newsandviews,Stockman_2018}. One of the pioneering works in quantum plasmonics is the demonstration that the optical resonances of localized surface plasmons can be strongly affected by the electronic confinement for metallic nanoparticle sizes of the order of 10~nm~\cite{nature_quantum}. In this case, the Drude model fails in describing the optical response of the electron gas, and a quantum treatment or non-local electromagnetic models~\cite{McMahon} must be considered. Size confinement also strongly affects the optical properties of polar materials~\cite{Ratchford}, and has required the use of computational models based on perturbative density functional theory~\cite{Paudel} or the use of a non-local dielectric response which goes beyond the Lorentz model to describe the material resonances~\cite{Gubbin}. Another fundamental quantum effect, tunneling, has been shown to strongly impact the properties of the plasmons beyond the classical treatement~\cite{savage, zhu, Tan}. 

In this work we investigate the frontier between classical and quantum plasmonics by studying the effect of a confining potential on volume plasmons. We observe that size confinement gives rise to several longitudinal modes with quantized wavevector at different energies, as a particle in a quantum well. We demonstrate that these modes do not obey the Lindhard formula, which holds for thin metallic films. Non-locality of volume plasmons in semiconductor layers clearly appears to be related to electronic size confinement, and is explained by using a quantum model which constructs the plasmon modes directly from the confined states of the constituent electrons.  
\begin{figure}[ht] 
\centering  
\includegraphics[width=0.7\columnwidth]{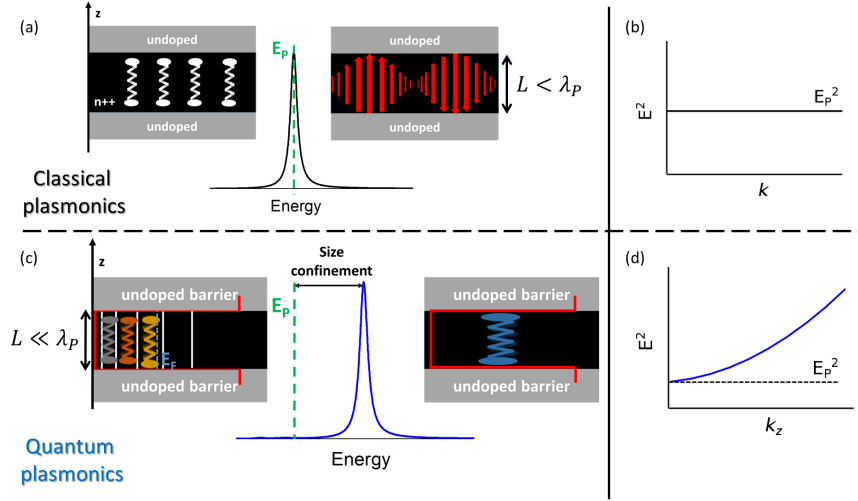}
\caption{(a) Sketch of a doped semiconductor layer embedded between two undoped layers. If the thickness $L$ is smaller than the plasma wavelength in the material, $\lambda_P=2 \pi c/\Omega_P$, it is possible to optically excite a Berreman mode at $E_P=\hbar \Omega_P$. (b) Dispersion of a volume plasmon in a classical model. (c) Effect of size confinement in the optical response of a semiconductor QW with three occupied electronic states. The optical response of the system is a collective resonance at a greater energy than $E_P$, due to size confinement contribution. (d) Longitudinal dispersion of a volume plasmon taking into account the material bandstructure and the Fermi distribution of electrons.}
\label{quantum_plasmonics}
\end{figure}

A sketch of the system is presented in fig.~\ref{quantum_plasmonics}(a). It is composed of a highly doped semiconductor layer, with thickness smaller than the plasma wavelength in the material. The oscillation of the free electrons in the doped layer excited by an incident electromagnetic field results in an optically active collective mode of the system: a plasmon confined in the doped layer with a dipole moment along the growth direction $z$. The absorption spectrum, sketched in fig.~\ref{quantum_plasmonics}(a), presents a single Lorentzian resonance, centered at the plasma energy, with a quality factor on the order of 10-20~\cite{Askenazi}. This resonance, called the Berreman mode~\cite{Ferrell,McAlister, Harbecke,Askenazi,newman}, can be simulated with the Drude model, which describes the isotropic permittivity of the doped semiconductor, and by also taking into account the finite thickness of the layer through standard electromagnetic simulations. This system is thus an example of {\em{classical plasmonics}}. Berreman modes, which have first been observed in thin metallic films~\cite{Ferrell, McAlister, Melnyk_PRL, Lindau1971_experiment_tonks_dattner}, have recently raised considerable attention because at the plasma frequency the real part of the dielectric permittivity is zero. For this reason, the Berreman mode is also referred to as an epsilon-near-zero (ENZ) mode~\cite{greffet, montano_2018}. Hyperbolic metamaterials obtained by alternating ENZ and dielectric layers have also been demonstrated~\cite{gmachl}. In the classical description, the plasma frequency is independent from the plasmon wavevector $\vec{k}$, as sketched in fig.~\ref{quantum_plasmonics}(b). As a consequence, the dielectric function is local, i.e. it only depends on the photon frequency and not on the wavevector. 

Quantum effects appear in semiconductors when the thickness of the layer is smaller than the de Broglie wavelength of electrons. In this case, size confinement along the $z$ direction gives rise to quantized energy levels. Figure~\ref{quantum_plasmonics}(c) sketches a doped semiconductor quantum well (QW), with three occupied confined states. In this case, there are three main optically active transitions, represented as three different sets of harmonic oscillators along $z$. Dipole - dipole interaction between these optically active transitions gives rise to a collective mode of the system, a confined plasmon, with an energy which is higher than the plasma energy of the electron gas~\cite{delteilPRL2012plasmons, Askenazi}, and higher than the energy of the individual electronic transitions. The Drude model thus fails to describe the collective optical properties of confined electrons, while quantum~\cite{delteilPRL2012plasmons, pegolottiPRB2014Multisubband_plasmons,montano_2018} or non-local semiclassical~\cite{AlpeggianiPRB2014Plasmon_semiclassic} models correctly reproduce the experimental absorption spectra. 

Optical experiments conducted on metal foils of few nm thickness~\cite{Lindau1971_experiment_tonks_dattner, Melnyk_PRL, melnykPRB1968_Tonks_dattner_theory} have shown the existence of higher order longitudinal modes, known as Tonks -- Dattner modes~\cite{Lindau_1971,raregas,ultracold}, whose energies are described by the Lindhard formula: $\Omega_{k_z}^2= \Omega_p^2 + \frac{3}{5} v_F^2 k_z^2$, where $\Omega_p$ is the bulk plasma frequency, $v_F$ the Fermi velocity and $k_z$ the $z$ component of the plasmon wavevector. This dispersion relation, sketched in fig.~\ref{quantum_plasmonics}(d), requires a quantum treatment of electrons in metals and a semiclassical treatment of light-matter interaction~\cite{wooten}.     
\begin{figure}[ht]
\centering  
\includegraphics[width=0.5\columnwidth]{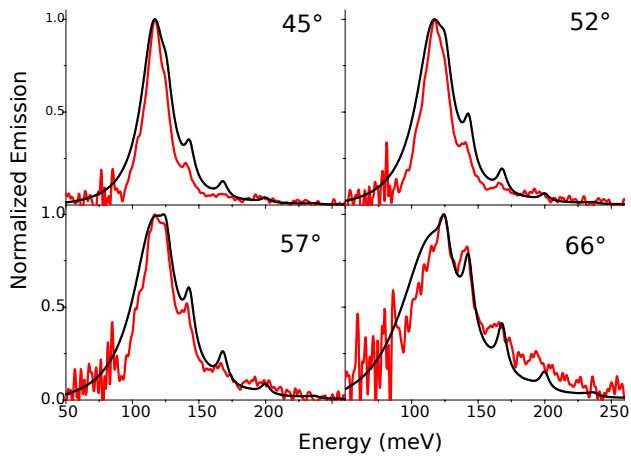}
\caption{Measured (red lines) and simulated (black lines) emission spectra under thermal excitation of the plasmons for different values of the internal angle of light propagation. The thermal emission spectra have been simulated by solving quantum Langevin equations in the input-output formalism~\cite{huppert_PRB2016}. The plasmon eigenmodes have been calculated from a full numerical diagonalization accounting for the finite barrier height as well as for nonparabolicity effects.} 
\label{spectres}
\end{figure}

In order to investigate the effect of a confining potential on the Berreman mode and on the higher order longitudinal plasmons, we have chosen to work within a semiconductor platform. Indeed, the quantum nature of the electrons is described straightforwardly in a semiconductor through the envelope function approximation that takes into account size confinement and tunneling, while in metals a quantum description requires numerical approaches.
Our sample is a doped 100~nm GaInAs/AlInAs layer with electronic density $N_v=7.5 \times 10^{18}$ cm$^{-3}$, embedded between two AlInAs barriers. We have performed angle resolved emission experiments under thermal excitation of the plasmons through the application of an in-plane current, as in ref.~\onlinecite{laurent2015}. Figure~\ref{spectres} shows normalized thermal emission spectra (in red) at different angles, measured at room temperature. Several peaks are clearly observed in the spectra. When increasing the angle, the main plasmon resonance, the Berreman mode, becomes broader and weaker due to the increase of the radiative decay rate of the plasmon~\cite{laurent2015}. Simultaneously, the higher energy resonances, the Tonks -- Dattner modes, become more and more visible. 

The energy position of all the plasmon modes and their angle dependent radiative broadening are very well reproduced by our quantum model (black lines in fig.~\ref{spectres}), which is based on the dipole representation of the light-matter interaction in the Coulomb gauge~\cite{todorovPRB2012plasmon_dipole_gauge, delteilPRL2012plasmons, pegolottiPRB2014Multisubband_plasmons,huppert_PRB2016}. In this model, the Hamiltonian of the electron gas in the QW is written in terms of the excitations between confined states as: 
\begin{equation}
H= \sum_{\alpha} \hslash \omega_{\alpha} b^\dag_{\alpha} b_{\alpha} \nonumber + \frac{e^2}{2\epsilon_0 \epsilon_s} \sum_{\alpha, \beta} S_{\alpha, \beta} \sqrt{\Delta N_{\alpha} \Delta N_{\beta}}\,  (b^\dag_{\alpha} + b_{\alpha})(b^\dag_{\beta} + b_{\beta}) \label{eq_H}
\end{equation}
Here the indices $\alpha, \beta$ run over all transitions between confined states in the QW; the operators $b^\dag_{\alpha}, b_{\alpha}$ are the creation and annihilation operators of the transition $\alpha$ of energy $\hbar \omega_\alpha$ and $\Delta N_\alpha$ 
the associated population difference; $\epsilon_s$ is the background dielectric constant. The coupling coefficients $S_{\alpha, \beta}$ are expressed as:
\begin{equation}
 S_{\alpha \, \beta}=\frac{1}{\hbar \omega_\alpha}\frac{1}{\hbar \omega_\beta}\left( \frac{\hbar^2}{2 m^*}\right)^2 \int_{-\infty}^{+\infty} {dz \, \xi_\alpha (z) \, \xi_\beta (z)}   
    \end{equation}
with $m^*$ the electron effective mass. The coupling coefficients between the electronic transitions $S_{\alpha, \beta}$ are proportional to the overlap between the a.c. microcurrent functions $\xi_\alpha$ describing the electronic transitions $\alpha=i \rightarrow i+j$ and defined as~\cite{todorovPRB2012plasmon_dipole_gauge}:
\begin{equation}
\xi_{\alpha} (z) \equiv \xi_{i \rightarrow i+j} (z) = \psi_i (z) \frac{\partial \psi_{i+j}(z)}{\partial z} -\psi_{i+j} (z) \frac{\partial \psi_{i}(z)}{\partial z},
\end{equation}
where $\psi_i$ is the envelope function of the confined electronic state with quantum number $i$. The integral of each microcurrent function is proportional to the optical dipole of the corresponding transition~\cite{todorovPRB2012plasmon_dipole_gauge,pegolottiPRB2014Multisubband_plasmons}. Note that in a square QW the transitions with a non-zero dipole are only those with odd $j$. 
\begin{figure}[t] 
\centering  
\includegraphics[width=0.7\columnwidth]{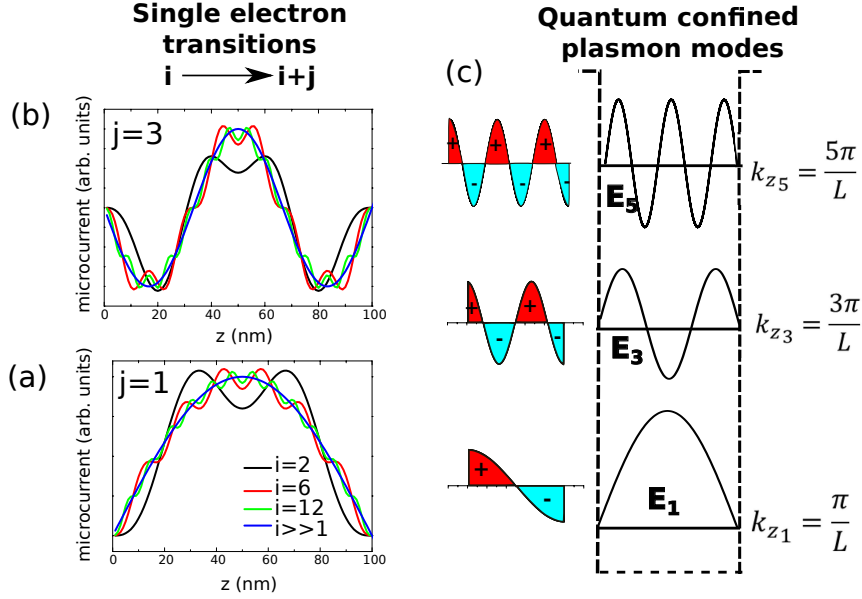}
\caption{(a), (b) Normalized microcurrents $\xi_{i \rightarrow i+j}(z)$ calculated for an infinite barrier QW for $j=1$ (a) and $j=3$ (b) and $i=2$ (black), $i=6$ (red), $i=12$ (green) and in the limit of $i \gg 1$ (blue). (c) Schematic representation of a square QW for the plasmon modes (dashed line). We have plotted in this effective potential the plasmon microcurrents (black continous lines) of the $j=1$, $j=3$ and $j=5$ modes, with $j$ the quantum number of the confined longitudinal wavevector. The charge density distributions are also plotted for each longitudinal mode.}
\label{micro_j}
\end{figure}

In the limit of infinite barriers, for a QW with thickness $L$, the microcurrent associated with the transition $i \rightarrow i+j$ writes: 
\begin{equation}
 \xi_{i \rightarrow i+j} (z) =\frac{\pi}{L^2} \left[ j \sin \left(\frac{2i+j}{L} \pi z\right) - (2i+j) \sin \left(\frac{j\pi z}{L} \right)   \right] \label{eq_micro}
\end{equation}
Examples of such functions for optically active transitions $i \rightarrow i+j$, with odd $j$, are plotted in fig.~\ref{micro_j} (a) ($j=1$) and (b) ($j=3$) for different values of $i$.  From both panels one can see that, for a fixed $j$, when increasing the value of the index $i$, the shape of the functions representing the microcurrents approaches $ \sin \left(\frac{j \, \pi z}{L} \right)$. As a consequence, in the limit $i \gg 1$, when a large number of subbands are occupied, the microcurrents $\left\{ \xi_{i \rightarrow i+j}\right\}_{i,j}$ are mutually orthogonal for different $j$. In this approximation, the coupling coefficients between microcurrents are given by: $S_{i \rightarrow i+j, i \rightarrow i+j'}=\dfrac{L}{2\pi^2 j \, j'}\delta_{j,j'}$. The matrix of the coupling coefficients is block - diagonal: all the electronic transitions with the same $j$ contribute to the same plasmon mode. In other words, the Hamiltonian \eqref{eq_H} can be independently diagonalized on the subspaces relative to transitions $i \rightarrow i+j$ for fixed $j$. For each subspace of index $j$, light couples with the plasmon corresponding to the highest frequency eigenmode. Its frequency $\Omega_j$ can be found, after Bogoliubov transformation within the associated subspace, by calculating the zeros of the following determinant~\cite{todorovPRB2012plasmon_dipole_gauge,pegolottiPRB2014Multisubband_plasmons,AlpeggianiPRB2014Plasmon_semiclassic}: 
\begin{eqnarray}
 \Delta_j (\omega) = 1- \frac{2e^2}{\hbar \epsilon_0 \epsilon_s} \sum_i \frac{S_{i \rightarrow i+j, i \rightarrow i+j} \  \Delta N_{i\rightarrow i+j} \   \omega_{i\rightarrow i+j}}{\omega^2-\omega_{i\rightarrow i+j}^2}= 1- \sum_{i=1}^{N_{occ}} \frac{{\omega_P}^2_{i,i+j}}{\omega^2-\omega_{i\rightarrow i+j}^2}.  \label{det_j}
\end{eqnarray}
with ${\omega_P}_{i,i+j}$ the plasma frequency associated with the transition $i \rightarrow i+j$.

Figure~\ref{micro_j}(c) presents the microcurrents associated with the confined plasmon modes issued from the subspaces of index $j=1,3,5$. These microcurrents present the same symmetry as those of the electronic transitions $i \rightarrow i+j$: they vary as $\sin{\left(\frac{j\pi z}{L} \right)}$ with a quantized wavevector $k_z=j \pi/L$. The oscillator strength of the plasmon mode $\Omega_j$ concentrates the interaction with light of all $i \rightarrow i+j$ transitions and it is given, for odd values of $j$, by $ \dfrac{4L\hbar N_v}{m^* \pi^2 j^2}$~\cite{pegolottiPRB2014Multisubband_plasmons, delteilPRL2012plasmons}. Figure~\ref{micro_j}(c) also presents the charge distributions oscillating at frequency $\Omega_j$, which are proportional to the derivative of the  microcurrents~\cite{pegolottiPRB2014Multisubband_plasmons}. We can clearly see that the fundamental mode $j=1$ corresponds to a dipole along $z$. This mode is thus the analogue of a Berreman mode in a metallic thin film. The modes with $j>1$ and $j$ odd are higher order longitudinal confined plasmons characterized by a quantized wavevector. They correspond to the Tonks - Dattner modes observed in the spectra in Fig.~\ref{spectres}. 

The zeros of the determinant, eq.~(\ref{det_j}), can be analytically calculated in two different regimes: the {\textit{metallic}} limit where a very large number of electronic states are occupied in an infinite potential well, and the {\textit{semiconductor}} limit which describes QWs with only few tens of electronic subbands occupied. In the {\textit{metallic}} limit, when the number of occupied subbands $N_{occ}$ is such that $j^2/N_{occ} \ll 1$, our model demonstrates the Lindhard formula (see supplementary information). However, in the QW that we have experimentally investigated we are far from the {\textit{metallic}} limit, as $N_{occ}=20$ and we observe higher order modes up to $j=9$, thus $j^2/N_{occ} \gg 1$. In our system, the energy separation between successive confined states is approximately constant, due to finite barrier and band non-parabolicity effects. In this case, we can introduce an average energy separation between the confined states, $E_0=\hbar \omega_0$, calculated as the conduction band offset $V_b$ divided by the total number of confined states in the QW, $N_{tot}$: $E_0=V_b/N_{tot}$. The zeros of the determinant, eq.~(\ref{det_j}) can be again calculated analytically, observing that $\Omega_P^2=\sum_{i=1}^{N_{occ}}{\omega_P}^2_{i,i+j}$~\cite{Askenazi}. The resulting plasmon mode frequencies are accurately described by:
\begin{equation}
\label{eq_plasmons}
\Omega_j^2=\Omega_p^2+ \omega_{0}^2 \, j^2 .
\end{equation}
From this equation it clearly appears that in the {\textit{semiconductor}} limit the observation of higher order plasmon modes is intimately related to the existence of an energy separation between the electronic states, $E_0$, induced by size confinement.

\begin{figure}[t] 
\centering  
\includegraphics[width=0.95\columnwidth]{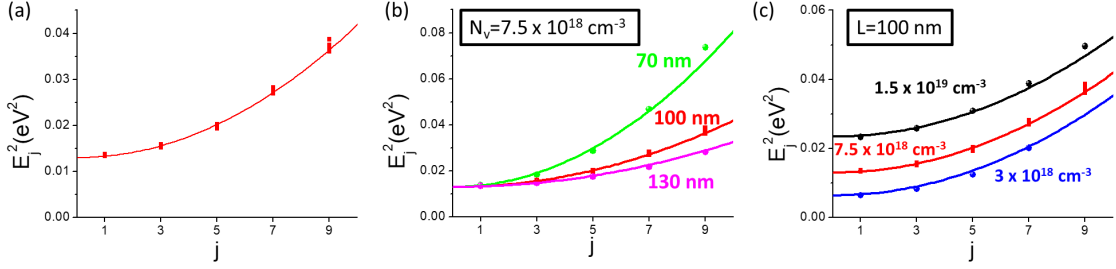}
\caption{(a) Squared energy of the plasmon modes extracted from the measured thermal emission spectra (red squares) plotted as a function of the quantum number $j$ of the corresponding confined plasmon. The line shows the result of eq.~\ref{eq_plasmons}. (b) and (c) Squared plasmon frequencies numerically calculated with our full quantum model (bullets), compared with the results of eq.~\ref{eq_plasmons} (lines). In panel (b) the electronic density has been set to $N_v=7.5 \times 10^{18}$~cm$^{-3}$, while the QW thickness is varied, while in panel (c) $L=100$~nm and the electronic density is varied. The experimental results are also reported for comparison (squares).} 
\label{verification_formula}
\end{figure}

Figure~\ref{verification_formula}(a) presents in red symbols the squared plasmon energies $E_j^2=(\hbar \Omega_j)^2$ extracted from 15 emission spectra measured at angles between 23$^\circ$ and 83$^\circ$ as a function of the index $j$ of the plasmon. The red line in Fig.~\ref{verification_formula} shows the calculated dispersion following eq.~\ref{eq_plasmons}, in excellent agreement with the experimental results. Note that this dispersion has been calculated with no free parameters, by using the calculated plasma energy $E_P=114$~meV, which takes into account band non-parabolicity~\cite{Askenazi}, and the confinement energy $E_{0}=17$~meV. From the excellent agreement between the experimental and the theoretical results we can infer that, notably, size confinement still plays a non-negligible role in a QW with $L=100$~nm, and it is at the origin of the observed non-locality effects~\cite{Gubbin}.  

In fig.~\ref{verification_formula}(b) and (c) we verify the validity of eq.~\ref{eq_plasmons} by varying the QW thickness for $N_v=7.5 \times 10^{18}$~cm$^{-3}$ (panel (b)) and the electronic density for $L=100$~nm (panel (c)). The squared plasmon frequencies are numerically calculated with our full quantum model (bullets) and compared with the results of eq.~\ref{eq_plasmons} (lines). Note that all the dispersions plotted in panel (b) have been calculated with the same value of the plasma energy $E_P=114$~meV, while those in panel (c) have been obtained with $E_0=17$~meV, independently on the electronic density.

\begin{figure} [t]
\centering  
\includegraphics[width=0.7\columnwidth]{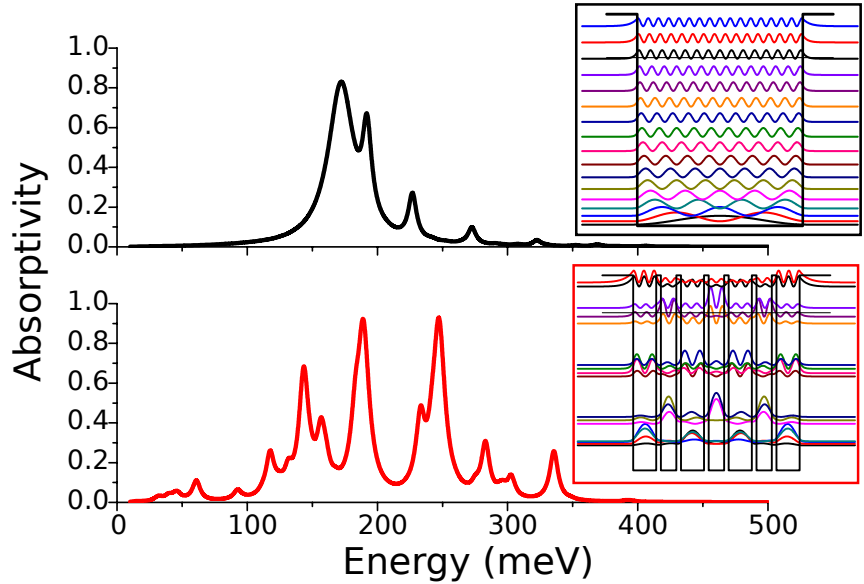}
\caption{Upper panel: Calculated absorptivity for a square GaInAs highly doped semiconductor layer, with thickness 54~nm and electronic density $N_v=2 \times 10^{19}$~cm$^{-3}$. Lower panel: Calculated absorptivity for a system of asymmetric tunnel coupled GaInAs/AlInAs quantum wells, with the same electronic density as the 54~nm structure. The conduction band profile and the square moduli of the wavefunctions for the two structures are presented as insets. The black horizontal lines indicate the Fermi energy.} 
\label{asymmetric}
\end{figure}

Having confirmed our microscopic approach in a square QW through the comparison between experimental and theoretical results, we discuss in the last part of this paper how electronic confinement and tunneling can be used as a degree of freedom to engineer the plasmonic resonances. The starting point is a GaInAs layer of 54~nm thickness, sandwiched between two AlInAs barriers. The electronic density per unit volume in the GaInAs layer is $N_v=2 \times 10^{19}$~cm$^{-3}$. Figure~\ref{asymmetric} presents the absorptivity spectrum simulated at $45^\circ$ (black line), showing the Tonks - Dattner resonances as previously discussed. The inset of the figure shows the corresponding band diagram, where we also plotted the square moduli of the electronic wavefunctions and the position of the Fermi energy (black horizontal line). We now insert in the GaInAs layer 6 identical AlInAs barriers, of 1.5~nm thickness, such that the structure is now composed of tunnel coupled asymmetric QWs as shown in the inset of fig.~\ref{asymmetric}. The electronic structure is profoundly modified by the presence of the barriers, resulting in the formation of several minibands. In particular, as the tunnel coupled QWs have different sizes, we now have optically active transitions between the states of the ground miniband and those of the second excited one (i.e. $j=2$, which are forbidden in a single QW)~\cite{Sirtori_asym}.
The electronic density per unit volume is kept equal to that of the single GaInAs layer, $N_v=2 \times 10^{19}$~cm$^{-3}$. The absorptivity spectrum including the collective effects is presented in red. This spectrum is completely different with respect to that obtained in the single layer case, proving that also collective excitations can be engineered by a judicious size confinement of the single electron states. Not only can many collective resonances be observed, but their oscillator strengths are distributed differently between them. In particular, the lowest energy collective mode is no longer the mode with the highest absorptivity. 

This work clarifies the link between the single electron wavefunctions and their collective response to the light. In that sense, it opens new degrees of freedom for engineering ad hoc plasmonic resonances in which the energy position and oscillator strength are determined by shaping single electron wavefunctions. Our theoretical approach is well suited for this goal, as it goes well beyond the Drude model, which is commonly used in semiconductor plasmonics~\cite{taliercio}, and the Lindhard formula, which applies to a free electron gas in a metallic thin film. Our model thus allows quantum engineering techniques to enter the field of semiconductor plasmonics. The ultimate goal is a complete three-dimensional shaping of the dielectric function, which fully exploits all the degrees of freedom offered by epitaxial growth and nanofabrication techniques in order to fabricate artificial ENZ materials for negative refraction~\cite{Hierro} and high harmonic generation~\cite{Yang} exploiting non-linear optical properties~\cite{Alam795}.

\begin{acknowledgments}
We acknowledge financial support from ERC (Grant ADEQUATE), Labex SEAM, and Agence Nationale de la Recherche (Grant No. ANR-19-CE30-0032-01).
\end{acknowledgments}

\bibliographystyle{apsrev4-1}
\bibliography{biblio}


\end{document}